\newcommand{\paperOption}{a4paper,onecolumn,12pt,twoside}
\newcommand{\geometryOption}{hscale=0.84,vscale=0.84,includehead,centering}
\newcommand{\hyperrefOption}{%
 pagebackref,pdfstartview=FitH,hidelinks%
}
\InputIfFileExists{option.aux}{%
 \typeout{Compiling options overridden!}%
}{}

\documentclass[\paperOption]{article}
\usepackage[\geometryOption]{geometry}
\usepackage[\hyperrefOption]{hyperref}

\usepackage{color}
\newcommand{\omitstyle}{\color[rgb]{0.25,0,0.75}}
\newcommand{\omitted}[1]{{\omitstyle#1}}
\InputIfFileExists{option.aux}{%
 \typeout{Compiling options activated!}%
}{}

\usepackage{clproject}
\input{urng.properties}

\usepackage{amsmath}
\usepackage{amssymb,amsthm}
\usepackage{mathenv}
\usepackage{cite}
\usepackage{graphics}
\usepackage[format=hang,font=small,labelfont=bf,margin=10pt]{caption}

\newtheorem{theorem}{Theorem}[section]
\newtheorem{corollary}[theorem]{Corollary}
\newtheorem{lemma}[theorem]{Lemma}
\newtheorem{proposition}[theorem]{Proposition}
\newtheorem{definition}[theorem]{Definition}
\newtheorem{example}[theorem]{Example}
\newtheorem{remark}[theorem]{Remark}

\newcommand{\eqdef}{:=}
\newcommand{\eqvar}[1]{\buildrel\mathrm{#1}\over=}
\newcommand{\eqdist}{\eqvar{d}}

\newcommand{\emptySet}{\varnothing}
\newcommand{\integer}{\mathbf{Z}}
\newcommand{\integerOf}[1]{\integer_{#1}}
\newcommand{\fromTo}[2]{[\![#1,#2]\!]}
\newcommand{\real}{\mathbf{R}}
\newcommand{\realOf}[1]{\real_{#1}}
\newcommand{\field}[1]{\mathbf{F}_{#1}}
\newcommand{\symmetricGroup}[1]{\mathrm{S}_{#1}}
\newcommand{\eConstant}{\mathrm{e}}

\newcommand{\setComplement}{\mathsf{c}}
\newcommand{\mapComposition}{\circ}
\newcommand{\floor}[1]{\left\lfloor#1\right\rfloor}
\newcommand{\ceil}[1]{\left\lceil#1\right\rceil}
\DeclareMathOperator{\sumOf}{\Sigma}
\DeclareMathOperator{\productOf}{\Pi}
\DeclareMathOperator{\statisticalDistance}{d}
\DeclareMathOperator{\expect}{E}
\DeclareMathOperator{\variance}{Var}
\DeclareMathOperator{\entropy}{H}
\DeclareMathOperator{\entropyRate}{\mathcal{H}}
\DeclareMathOperator{\infEntropy}{\underline{\entropy}}
\DeclareMathOperator{\infEntropyRate}{\underline{\entropyRate}}
\DeclareMathOperator*{\pliminf}{p-lim\,inf}

\DeclareMathOperator{\smallO}{o}
\DeclareMathOperator{\id}{id}

\DeclareMathOperator{\rate}{\mathsf{R}}
\DeclareMathOperator{\region}{\mathcal{R}}
\newcommand{\uniformDistribution}[1]{\mathrm{U}_{#1}}

\newcommand{\affineRandomBinning}[1]{\mathsf{A}_{#1}}
\newcommand{\linearRandomBinning}[1]{\mathsf{L}_{#1}}

\newcommand{\terminology}[1]{\emph{#1}}
\newcommand{\informationTheoretic}{\mathrm{i}}

\newcommand{\theTitle}{%
 Separate Random Number Generation from Correlated Sources%
}

\title{%
 \theTitle%
 \footnotetext{Version: \exactProjectVersion\ (no.~\releaseNumber).}%
 \footnotetext{
  This file was generated on \today\ by \texCompilerName{} with format
  \fmtname\ \fmtversion.
 }%
 \footnotetext{Paper options: $\mathrm{\paperOption}$.}%
 \footnotetext{\textsf{geometry} options: $\mathrm{\geometryOption}$.}%
 \footnotetext{\textsf{hyperref} options: $\mathrm{\hyperrefOption}$.}%
}
\author{%
 Shengtian Yang%
 \thanks{%
  S. Yang is with the School of Information and Electronic Engineering,
  Zhejiang Gongshang University, Hangzhou 310018, China (e-mail:
  \href{mailto:yangst@codlab.net}{\texttt{yangst@codlab.net}}).
 }%
}
\date{}

\begin{document}

\maketitle
\thispagestyle{myheadings}
\pagestyle{myheadings}

\begin{abstract}
This work studies the problem of separate random number generation
from correlated general sources with side information at the tester
under the criterion of statistical distance.
Tight one-shot lower and upper performance bounds are obtained using
the random-bin approach.
A refined analysis is further performed for two important random-bin
maps.
One is the pure-random-bin map that is uniformly distributed over the
set of all maps (with the same domain and codomain).
The other is the equal-random-bin map that is uniformly distributed
over the set of all surjective maps that induce an equal or
quasi-equal partition of the domain.
Both of them are proved to have a doubly-exponential concentration of
the performance of their sample maps.
As an application, an open and transparent lottery scheme, using a
random number generator on a public data source, is proposed to solve
the social problem of scarce resource allocation.
The core of the proposed framework of lottery algorithms is a
permutation, a good rateless randomness extractor, whose existence is
confirmed by the theoretical performance of equal-random-bin maps.
This extractor, together with other important details of the scheme,
ensures that the lottery scheme is immune to all kinds of fraud under
some reasonable assumptions.
\end{abstract}

\begin{quotation}
\end{quotation}

\section{Introduction}\label{introduction}

The problem of random number generation, namely extracting randomness
from a random source, may date back to von Neumann
\cite{vonNeumann195100},
who designed a simple algorithm for simulating a fair coin by using a
biased coin with unknown probability.
So far, there has been a large body of research in this area, from
theory to practice, but basically, all research can be classified into
two categories.

The first category takes a model-based approach, assuming that the
probability distribution of source is known or belongs to a family of
probability distributions, e.g.,
\cite{vembu_generating_1995,han199703,visweswariah199803,han200300,
 zhou201204,seroussi_optimal_2015}
and the references therein.
The main issues in this category are the estimation of model
parameters and the extraction of randomness from a source with known
statistics.
Traditional methods of lossless data compression
with necessary modifications work well in this case.

In contrast, the second category of research, which is popular in
computer science, does not make any assumption on the source, except
that the source must have enough randomness to be extracted, e.g.,
\cite{nisan199902,guruswami200906}
and the references therein.
A so-called min-entropy quantity, usually much less than the entropy
of a source, is used to measure the amount of randomness of a source.
The key issue in this category is designing a seeded (randomness)
extractor that can extract a specified number of random bits from any
data having enough randomness in terms of min-entropy.
Its underlying theory is extensively related to universal hashing
\cite{cater197904},
or similarly, random bins, a technique widely used in distributed
lossless source coding (i.e., Slepian-Wolf coding or separate coding
of correlated sources) \cite{cover197503}.

By the classification above, we show a clear duality between random
number generation and lossless source coding.
This duality is however not perfect in that the performance bound of
extractors is characterized by min-entropy
\cite{radhakrishnan200001}
whereas the achievable rate region of Slepian-Wolf coding is
characterized by entropy
\cite{slepian197307}.
Intuitively, since these two areas use the same principle of random
bins, one may conjecture that there is a perfect duality between them.
Does such a duality really exist?
The answer is positive and is in fact partly known in recent research
of random number generation from correlated sources (see
Remarks~\ref{directPartRemark} and \ref{conversePartRemark}), though
not explicitly stated.
In this paper, we will investigate this problem in depth.
The contribution of this paper can be summarized as follows:

1) One-shot lower and upper bounds for the performance of separate
random number generation from correlated sources with side information
at the tester are derived.
They are tight enough for the first- and second-order performance
analysis.
The upper bound even holds true for all $1$-random-bin maps, a
generalization of the classical random binning.

2) A phenomenon of doubly-exponential concentration of performance is
proved for two kinds of random-bin maps, one of which is the
equal-random-bin map, an important ensemble that has not received much
attention in information theory.
This result deepens our understanding of random number generation and
gives us some useful hints on the design of random number generator.

3) A lottery scheme, using a random number generator on a public data
source, is proposed to solve the social problem of scarce resource
allocation.
The core of the proposed framework of lottery algorithm is a
permutation, a good rateless randomness extractor, whose existence is
confirmed by the theoretical performance of equal-random-bin maps.

The rest of the paper is organized as follows:
Section~\ref{formulation} formulates the problem of separate random
number generation from correlated sources with side information at the
tester.
One-shot performance bounds as well as the achievable rate region of
this problem is presented in Section~\ref{basicResults}.
A refined analysis is performed in Section~\ref{refinedAnalysis} for
two important extractor ensembles, which turn out to have a sharp
concentration of the performance of their individual extractors.
In Section~\ref{lottery}, we propose an open and transparent lottery
scheme for resource allocation.

We close this section with some notations used throughout the paper.
The ring of integers, the field of real numbers, and the finite field
of order $q$ are denoted by $\integer$, $\real$, and $\field{q}$,
respectively.
A subset of $\integer$ (and similarly for $\real$) is usually denoted
by $\integerOf{A}\eqdef \integer\cap A$ for some set $A$.
Then the set of integers from $1$ to $n$ is denoted
$\integerOf{[1,n]}$, and sometimes, if $A=[0,+\infty)$ for example, we
simply write $\integerOf{\ge 0}$ in place of
$\integerOf{[0,+\infty)}$.
Since $\integerOf{[m,n]}$ will be frequently used, we further define
its shorthand $\fromTo{m}{n}$, where $m$ and $n$ may not be integers.

A sequence (or more generally, an indexed family) is a map $x$ from an
index set $I$ into a value collection $\mathcal{X}$.
Usually, we denote it by $x=(x_i)_{i\in I}$ (or $(x_i)_{i=1}^\infty$
if $I=\integerOf{>0}$), where $x_i$ is the shorthand of $x(i)$.
The index $i$ can also be written as the superscript of $x$, but we
parenthesize it so that it is not confused with the power operation.
When, for example, $I=\integerOf{>0}^3$, we may further use the
notation such as $x=(x_{i,j}^{(k)})_{i,j,k\in\integerOf{>0}}$.
A partial sequence of $x$ is simply the restriction $x|_A$ of map $x$
to some $A\subseteq I$, and we write $x_A=(x_i)_{i\in A}$ as
shorthand, so that $x$ becomes the further shorthand of $x_I$.
Then, for example, for $x=(x_i^{(j)})_{i,j\in\integerOf{>0}}$, the
notations $x_{\integerOf{>0}}$, $x^{(\integerOf{>0})}$,
$x_{\integerOf{>0}}^{(\integerOf{>0})}$ all refer to the same sequence
$x$, and $x^{(j)}$ or $x_{\integerOf{>0}}^{(j)}$ refers to the partial
sequence $(x_i^{(j)})_{i=1}^\infty$.
For convenience of notation, we also define
$\sumOf(x)\eqdef \sum_{i\in I} x_i$ and
$\productOf(x)\eqdef \prod_{i\in I} x_i$ whenever these operations
make sense for $x$.
Then for a family $\mathcal{X}=\mathcal{X}_I$ of alphabets, we write
$\productOf(\mathcal{X})$ in place of the cartesian product
$\prod_{i\in I} \mathcal{X}_i$.

When performing probabilistic analysis, all objects of study are
related to a basic probability space $(\Omega,\mathfrak{A},P)$ with
$\mathfrak{A}$ a $\sigma$-algebra in $\Omega$ and $P$ a probability
measure on $(\Omega,\mathfrak{A})$.
A random element in a measurable space $(\Omega',\mathfrak{B})$ is a
measurable mapping from $\Omega$ into $\Omega'$.
A random element uniformly distributed over $\Omega'$ is denoted by
$\uniformDistribution{\Omega'}$.
For random elements $X$ and $Y$ in $\Omega'$, the
\terminology{statistical distance} between $P_X$ and $P_Y$ is
\[
\statisticalDistance(P_X,P_Y)
\eqdef \sup_{B\in\mathfrak{B}} |P_X(B)-P_Y(B)|.
\]
When $\Omega'$ is at most countable and $\mathfrak{B}$ is the power
set of $\Omega'$, we have
\[
\statisticalDistance(P_X,P_Y)
= \frac{1}{2}\sum_{\omega'\in\Omega'} |P_X(\omega')-P_Y(\omega')|.
\]
If $\statisticalDistance(P_X,P_Y)=0$, then $X$ and $Y$ have the same
probability distribution and we write $X\eqdist Y$.
A mixture distribution of $P_X$ and $P_Y$, usually denoted
$\lambda P_X+(1-\lambda)P_Y$, is a convex combination of $P_X$ and
$P_Y$ with some nonnegative weights $\lambda$ and $1-\lambda$.

\section{Problem Formulation}\label{formulation}

Suppose that there are $m+1$ correlated sources, indexed from $0$ to
$m$.
With each source of positive index, there is associated a separate
(seedless) extractor.
Roughly speaking, we hope that the output data of all extractors are
not only close to uniform but also almost independent of the source of
index zero.
Under this condition, we wonder how much randomness we can extract
from the correlated sources.

The formal definition is given as follows, using the
information-spectrum approach \cite{han200300}:

\begin{definition}[Correlated sources]\label{problemFormulation.1}
Let $\mathcal{X} = (\mathcal{X}_i)_{i\in\fromTo{0}{m}}$ denote an
$(m+1)$-tuple of alphabets, where $\mathcal{X}_0$ is at most countable
and all the other $\mathcal{X}_i$'s are finite.
An $(m+1)$-tuple of correlated (general) sources is a sequence
$X=(X_{\fromTo{0}{m}}^{(n)})_{n=1}^\infty$ of random elements
\[
X_{\fromTo{0}{m}}^{(n)}
= (X_{i,j}^{(n)})_{i\in\fromTo{0}{m},j\in\fromTo{1}{n}},
\]
where $i$ is the source index, $j$ is the time index, and each
$X_{i,j}^{(n)}$ is a random element in $\mathcal{X}_i$.
When $X$ is stationary and memoryless, it can simply be identified
with the random element $(X_0^{(1)},X_1^{(1)},\ldots,X_m^{(1)})$ (or
$(X_{0,j}^{(n)},X_{1,j}^{(n)},\ldots,X_{m,j}^{(n)})$ for any
$j\in\fromTo{1}{n}$) in
$\mathcal{X}_0\times\mathcal{X}_1\times\cdots\times\mathcal{X}_m$.
\end{definition}

\begin{definition}[Extractors]\label{problemFormulation.2}
Let
$\varphi^{(n)} = (\varphi_i^{(n)}:
 \mathcal{X}_i^n\to\mathcal{Y}_i^{(n)})_{i\in\fromTo{1}{m}}$
be an $m$-tuple of (seedless) extractors,%
\footnote{
 There are two kinds of randomness extractors, seeded and seedless,
 but there is no substantial difference between them, since a general
 source combined with a random seed is again a general source.
}
with
$\mathcal{Y}^{(n)} = (\mathcal{Y}_{i}^{(n)})_{i\in\fromTo{1}{m}}$
denoting the $m$-tuple of output alphabets.
The rate of each $\varphi_i^{(n)}$ is defined as
\[
\rate(\varphi_i^{(n)})
\eqdef \frac{1}{n}\ln|\mathcal{Y}_i^{(n)}|.
\]
In the sequel, most of our analysis will be performed on some ensemble
of extractors endowed with some probability measure, viz., a random
extractor.
For convenience, we assume from now on that \emph{every random
extractor is independent of all the other random objects in question}.
\end{definition}

\begin{definition}[Achievable rate region]\label{problemFormulation.3}
An $m$-tuple $R = (R_i)_{i\in\fromTo{1}{m}}$ of rates is said to be
achievable if there exists an $m$-tuple $\varphi^{(n)}$ of extractors
such that
\[
\liminf_{n\to\infty} \rate(\varphi_i^{(n)})
\ge R_i
\quad\text{for all $i\in\fromTo{1}{m}$}
\]
and
\[
\lim_{n\to\infty} \statisticalDistance(
 P_{X_0^{(n)}(\varphi_i^{(n)}(X_i^{(n)}))_{i\in\fromTo{1}{m}}},
 P_{X_0^{(n)}}P_{\uniformDistribution{\productOf(\mathcal{Y}^{(n)})}})
= 0.
\]
Then the achievable rate region of $X$ is
\[
\region(X)
\eqdef \{R\in\realOf{\ge 0}^m:\text{$R$ is achievable}\}.
\]
\end{definition}

\section{One-Shot Bounds and Achievable Rate Region}
\label{basicResults}

In this section, we will derive one-shot lower and upper performance
bounds as well as the achievable rate region $\region(X)$.
Since random bins are our main tool for constructing extractors, we
first give an exact definition of random bins, which is more general
than needed.

\begin{definition}\label{randomBin}
Let $F$ be a random map from $A$ to $B$, with $A$ and $B$ both finite.
The map $F$ is said to be a
\terminology{$(Q,\gamma)$-random-bin map} for some probability
distribution $Q$ on $B$ and some $\gamma\ge 0$ if for distinct
$x,y\in A$ and $z\in B$,
\[
P\{F(x)=z\} = Q(z)
\]
and
\begin{equation}
P\{F(x)=z,F(y)=z\} \le \gamma Q(z)^2.\label{randomBinCondition.2}
\end{equation}
When $Q$ is the uniform distribution on $B$, we simply say that $F$ is
a $\gamma$-random-bin map.
\end{definition}

Below we provide some examples of $\gamma$-random-bin maps, which will
be needed later on.

\begin{example}[Pure random binning]\label{prb}
A \terminology{pure-random-bin map} $F$ is a random map uniformly
distributed over the set of all maps of $A$ into $B$.
It is clear that $F$ is a $1$-random-bin map.
\end{example}

\begin{example}[Equal random binning]\label{erb}
A map $f$ of $A$ onto $B$ is said to be an \terminology{equal-bin map}
if
\[
\floor{\frac{|A|}{|B|}}
\le |f^{-1}(z)|
\le \ceil{\frac{|A|}{|B|}}
\]
for all $z\in B$.
An \terminology{equal-random-bin map} $F$ is a random map uniformly
distributed over the set of all equal-bin maps from $A$ onto $B$.
It is clear that
$F\eqdist \uniformDistribution{\symmetricGroup{B}} \mapComposition f
 \mapComposition \uniformDistribution{\symmetricGroup{A}}$,
where $\symmetricGroup{A}$ denotes the symmetric group of all
permutations of $A$ and $f$ is an arbitrary fixed equal-bin map from
$A$ onto $B$.
It is easy to figure out that for distinct $x,y\in A$ and $z\in B$,
\[
P\{F(x)=z\} = \frac{1}{|B|}
\]
and
\[
P\{F(x)=z,F(y)=z\}
= \frac{(|A|-r)(|A|-|B|+r)}{|A|(|A|-1)|B|^2}
\le \frac{1}{|B|^2},
\]
where $r=|A|\bmod|B|$.
Therefore, $F$ is a $1$-random-bin map.
\end{example}

\begin{example}[Affine random binning]
Suppose that $A=\field{q}^m$ and $B=\field{q}^n$.
Let $C$ be an $(m,n,k)$ maximum-rank-distance (MRD) code over
$\field{q}$ with $1\le k\le \min\{m,n\}$
\cite{delsarte197800,gabidulin198501}.
An \terminology{affine-random-bin map}
$\affineRandomBinning{C}$ is a random affine map from
$\field{q}^m$ to $\field{q}^n$ given by
$v\mapsto v\uniformDistribution{C}+\uniformDistribution{\field{q}^n}$
(where both $\field{q}^m$ and $\field{q}^n$ are regarded as row-vector
spaces).
By \cite[Theorem~2.5 and Proposition~5]{yang201205},
$\affineRandomBinning{C}$ is a $1$-random-bin map.
In particular, the set $\field{q}^{m\times n}$ of all $m\times n$
matrices over $\field{q}$ is an $(m,n,\min\{m,n\})$ MRD code, so that
$\affineRandomBinning{\field{q}^{m\times n}}$ is a $1$-random-bin map.
\end{example}

\begin{example}[Binary linear random binning]\label{lrb}
Suppose that $A=\field{2}^m$ and $B=\field{2}^n$.
Let $C$ be an $(m,n,k)$ MRD code over $\field{2}$ with
$2\le k\le \min\{m,n\}$.
A \terminology{binary linear-random-bin map} $\linearRandomBinning{C}$
is a random linear map from $\field{2}^m$ to $\field{2}^n$ given by
$v\mapsto v \uniformDistribution{C}$.
By \cite[Theorem~3.6]{yang201205},
$\linearRandomBinning{C}|_{\{0\}^\setComplement}$ is a $1$-random-bin
map.
In particular,
$\linearRandomBinning{\field{2}^{m\times n}}|_{\{0\}^\setComplement}$
is a $1$-random-bin map when $m,n\ge 2$.
\end{example}

For $(Q,\gamma)$-random-bin maps, we have the following important
property:

\begin{theorem}\label{randomBinProperty.1}
Let $\mathcal{X}=\mathcal{X}_{\fromTo{0}{m}}$ and
$\mathcal{Y}=\mathcal{Y}_{\fromTo{1}{m}}$ be two families of alphabets
with $\mathcal{X}_0$ at most countable and all the other alphabets
finite.
Let $X=(X_i)_{i\in\fromTo{0}{m}}$ be a random element in
$\productOf(\mathcal{X})$ and $Y=(Y_i)_{i\in\fromTo{1}{m}}$ a random
element in $\productOf(\mathcal{Y})$ with all $Y_i$'s mutually
independent (and also independent of any other random object).
Let $F=(F_i:\mathcal{X}_i\to\mathcal{Y}_i)_{i\in\fromTo{1}{m}}$ be an
$m$-tuple of random maps such that each $F_i$ is an independent
$(P_{Y_i},\gamma_i)$-random-bin map, where
$\gamma=(\gamma_i)_{i\in\fromTo{1}{m}}$ is an $m$-tuple of real
numbers.
Then, for $a\in \mathcal{X}_0$ and $y\in\productOf(\mathcal{Y})$,
\[
\expect[\mu_X(F,a,y)]
= P_{X_0}(a)P_Y(y)
\]
and
\[
\variance(\mu_X(F,a,y))
\le (\productOf(\gamma)-1)P_{X_0}(a)^2P_Y(y)^2
 + \sum_{\substack{S\subseteq\fromTo{1}{m}\\ S\ne\emptySet}}
 \productOf(\gamma_{S^\setComplement})P_Y(y)
 P_{Y_{S^\setComplement}}(y_{S^\setComplement}) \nu_X(a,S),
\]
where
\begin{eqnarray*}
\mu_X(F,a,y)
&\eqdef &P[X_0=a,(F_i(X_i)=y_i)_{i\in\fromTo{1}{m}}\mid F],\\
\nu_X(a,S)
&\eqdef &\sum_{u\in\productOf(\mathcal{X}_S)} P_{X_0X_S}(a,u)^2,
\end{eqnarray*}
and $S^\setComplement = \fromTo{1}{m}\setminus S$.%
\footnote{
 This convention will be used in the sequel for subsets of index set
 $\fromTo{1}{m}$ of correlated sources.
}
\end{theorem}

\begin{proof}
Using the property of conditional expectation, we have
\begin{eqnarray*}
\expect[\mu_X(F,a,y)]
&= &P\{X_0=a,(F_i(X_i)=y_i)_{i\in\fromTo{1}{m}}\}\\
&= &\expect[P[X_0=a,(F_i(X_i)=y_i)_{i\in\fromTo{1}{m}}
 \mid X]]\\
&= &\expect\left[1\{X_0=a\}\prod_{i=1}^m P_{Y_i}(y_i)\right]
= P_{X_0}(a)P_{Y}(y).
\end{eqnarray*}
Since
$\variance(\mu_X(F,a,y)) =
 \expect[\mu_X(F,a,y)^2]-(\expect[\mu_X(F,a,y)])^2$,
it suffices to calculate the first term at the right-hand side.
We have
\begin{eqnarray}
& &\expect[\mu_X(F,a,y)^2]\nonumber\\
& &= \expect\left[\left(\sum_{x\in\mathcal{X}}
 P_X(x)1\{x_0=a,(F_i(x_i)=y_i)_{i\in\fromTo{1}{m}}\}
 \right)^2\right]\nonumber\\
& &\le \sum_{S\subseteq\fromTo{1}{m}}
 \sum_{u\in\productOf(\mathcal{X}_S)}
 \sum_{\substack{v,w\in\productOf(\mathcal{X}_{S^\setComplement})\\
  (v_i\ne w_i)_{i\in S^\setComplement}}}
 P_{X_0X_SX_{S^\setComplement}}(a,u,v)
 P_{X_0X_SX_{S^\setComplement}}(a,u,w) P_Y(y)
 \productOf(\gamma_{S^\setComplement})
 P_{Y_{S^\setComplement}}(y_{S^\setComplement})
 \label{eq:randomBinProperty.1.key1}\\
& &\le \sum_{S\subseteq\fromTo{1}{m}}
 \productOf(\gamma_{S^\setComplement}) P_Y(y)
 P_{Y_{S^\setComplement}}(y_{S^\setComplement})
 \sum_{u\in\productOf(\mathcal{X}_S)}
 \left(\sum_{v\in\productOf(\mathcal{X}_{S^\setComplement})}
 P_{X_0X_SX_{S^\setComplement}}(a,u,v)\right)^2\nonumber\\
& &= \sum_{S\subseteq\fromTo{1}{m}}
 \productOf(\gamma_{S^\setComplement})P_Y(y)
 P_{Y_{S^\setComplement}}(y_{S^\setComplement})
 \sum_{u\in\productOf(\mathcal{X}_S)} P_{X_0X_S}(a,u)^2,\nonumber
\end{eqnarray}
where \eqref{eq:randomBinProperty.1.key1} follows from the expansion
of the product and the inequality
\begin{eqnarray*}
& &\expect[
 1\{x_0=a,F_S(u)=y_S,F_{S^\setComplement}(v)=y_{S^\setComplement}\}
 1\{x_0=a,F_S(u)=y_S,F_{S^\setComplement}(w)=y_{S^\setComplement}\}]\\
& &= \expect[1\{x_0=a\}1\{F_S(u)=y_S\}
 1\{F_{S^\setComplement}(v)=y_{S^\setComplement},
 F_{S^\setComplement}(w)=y_{S^\setComplement}\}]\\
& &\le 1\{x_0=a\}P_{Y_S}(y_S)\productOf(\gamma_{S^\setComplement})
 [P_{Y_{S^\setComplement}}(y_{S^\setComplement})]^2
= 1\{x_0=a\}P_{Y}(y)\productOf(\gamma_{S^\setComplement})
 P_{Y_{S^\setComplement}}(y_{S^\setComplement}),
\end{eqnarray*}
so that
\[
\variance(\mu_X(F,a,y))
\le (\productOf(\gamma)-1)P_{X_0}(a)^2P_Y(y)^2
 + \sum_{\substack{S\subseteq\fromTo{1}{m}\\ S\ne\emptySet}}
 \productOf(\gamma_{S^\setComplement})P_Y(y)
 P_{Y_{S^\setComplement}}(y_{S^\setComplement}) \nu_X(a,S).\qedhere
\]
\end{proof}

A useful consequence follows immediately from
Theorem~\ref{randomBinProperty.1}.

\begin{corollary}\label{randomBinProperty.2}
Under the same condition as Theorem~\ref{randomBinProperty.1},
\[
\expect[d(X,Y\mid F)]
\le \frac{1}{2}\sum_{a\in\mathcal{X}_0,y\in\mathcal{Y}}
 \left[(\productOf(\gamma)-1)P_{X_0}(a)^2P_Y(y)^2
 + \sum_{\substack{S\subseteq\fromTo{1}{m}\\S\ne\emptySet}}
 \productOf(\gamma_{S^\setComplement})P_Y(y)
 P_{Y_{S^\setComplement}}(y_{S^\setComplement}) \nu_X(a,S)
 \right]^{\frac{1}{2}}.
\]
where
$d(X,Y\mid F)\eqdef \statisticalDistance(
 P_{X_0(F_i(X_i))_{i\in\fromTo{1}{m}}}[\cdot\mid F],P_{X_0}P_{Y})$
is a $\sigma(F)$-measurable random variable.
\end{corollary}

\begin{proof}
It is clear that
\begin{eqnarray}
\expect[d(X,Y\mid F)]
&= &\expect\left[
 \frac{1}{2}\sum_{a\in\mathcal{X}_0,y\in\mathcal{Y}}
 |\mu_X(F,a,y)-P_{X_0}(a)P_Y(y)|\right]\nonumber\\
&= &\frac{1}{2}\sum_{a\in\mathcal{X}_0,y\in\mathcal{Y}}
 \expect[|\mu_X(F,a,y)-P_{X_0}(a)P_Y(y)|]\nonumber\\
&\le &\frac{1}{2}\sum_{a\in\mathcal{X}_0,y\in\mathcal{Y}}
 \sqrt{\variance(\mu_X(F,a,y))},\label{eq:randomBinProperty.2.key1}
\end{eqnarray}
which combined with Theorem~\ref{randomBinProperty.1} establishes the
corollary, where \eqref{eq:randomBinProperty.2.key1} follows from
Jensen's inequality.
\end{proof}

We are now in the position to derive tight one-shot upper and lower
bounds, which are the crux of the proof of the achievable rate region
(Theorem~\ref{regionCharacterization}) and are useful in a refined
analysis for special random-bin maps (Section~\ref{refinedAnalysis})
or in the second-order performance analysis.

\begin{lemma}\label{directPart}
Let $X=X_{\fromTo{0}{m}}$ be an $(m+1)$-tuple of random elements in
$\productOf(\mathcal{X})$ with
$\mathcal{X}=\mathcal{X}_{\fromTo{0}{m}}$.
Let
$\Phi=\{\Phi_i:\mathcal{X}_i\to\mathcal{Y}_i\}_{i\in\fromTo{1}{m}}$ be
an $m$-tuple of $1$-random-bin extractors.
Then for $r>1$,
\[
\expect[d(X\mid\Phi)]
\le P\left\{T_X(X)\notin A_r(\mathcal{Y})\right\} + \frac{\sqrt{2^m-1}}{2}r^{-1/2},
\]
where
\begin{eqnarray}
d(X\mid\Phi)
&\eqdef
&d(X,\uniformDistribution{\productOf(\mathcal{Y})}\mid\Phi),
\label{eq:DeltaDistance}\\
T_X(x)
&\eqdef &(T_X^{(S)}(x))_{\emptySet\ne S\subseteq\fromTo{1}{m}}
= \left(\frac{1}{P_{X_S\mid X_0}(x_S\mid x_0)}\right)_{\emptySet\ne S\subseteq\fromTo{1}{m}},
\label{eq:Statistics}\\
A_r(\mathcal{Y})
&\eqdef &\prod_{\emptySet\ne S\subseteq\fromTo{1}{m}}
 I_{r|\productOf(\mathcal{Y}_S)|},
\label{eq:TypicalSets}\\
I_t
&\eqdef &(t,+\infty).
\label{eq:Interval}
\end{eqnarray}
\end{lemma}

\begin{proof}
Let $\lambda=P(G)$ with $G=\{\omega:T_X(X)\in A_r(\mathcal{Y})\}$.
Since the lemma holds trivially for $\lambda=0$, we suppose that
$\lambda>0$.
Then
\[
P_X= \lambda P_V+(1-\lambda) P_W,
\]
where $P_V=P_{X\mid 1_G=1}$ and $P_W=P_{X\mid 1_G=0}$ with $1_G$
denoting the indicator function of $G$.
In particular, $P_V(x)\le P_X(x)/\lambda$ for all
$x\in\productOf(\mathcal{X})$.
It is clear that
\begin{eqnarray*}
d(X\mid\Phi)
&\le &\lambda d(V\mid\Phi)
 + (1-\lambda) d(W\mid\Phi)\\
&\le &\lambda d(V\mid\Phi) + 1-\lambda,
\end{eqnarray*}
where the first inequality is due to
Proposition~\ref{statisticalDistanceProperty.1}.
From Corollary~\ref{randomBinProperty.2}, it follows that
\begin{eqnarray*}
\expect[d(V\mid\Phi)]
&\le &\frac{1}{2}\sum_{a\in\mathcal{X}_0,y\in\mathcal{Y}}
 \left(\sum_{\emptySet\ne S\subseteq\fromTo{1}{m}}
 \frac{\nu_V(a,S)}{|\productOf(\mathcal{Y})|
  |\productOf(\mathcal{Y}_{S^\setComplement})|}
 \right)^{\frac{1}{2}}\\
&= &\frac{1}{2}\sum_{a\in\mathcal{X}_0}
 \left(\sum_{\emptySet\ne S\subseteq\fromTo{1}{m}}
 |\productOf(\mathcal{Y}_S)| \nu_V(a,S) \right)^{\frac{1}{2}}.
\end{eqnarray*}
Since $V\in X(G)$ almost surely, we have
$T_X^{(S)}(V) > r|\productOf(\mathcal{Y}_S)|$ almost surely, so that
$P_{X_S\mid X_0}(V_S\mid V_0)<p=1/(r|\productOf(\mathcal{Y}_S)|)$
almost surely, and hence
\begin{eqnarray*}
\nu_V(a,S)
&= &\sum_{u:P_{X_S\mid X_0}(u\mid a)<p} P_{V_0V_S}(a,u)^2\\
&\le &\sum_{u:P_{X_S\mid X_0}(u\mid a)<p}
 \left(\sum_{v\in\productOf(\mathcal{X}_{S^\setComplement})}
 \frac{P_{X_0X_SX_{S^\setComplement}}(a,u,v)}{\lambda}\right)^2\\
&= &\lambda^{-2} \sum_{u:P_{X_S\mid X_0}(u\mid a)<p}
 P_{X_0X_S}(a,u)^2\\
&= &\lambda^{-2} P_{X_0}(a)^2 \sum_{u:P_{X_S\mid X_0}(u\mid a)<p}
 P_{X_S\mid X_0}(u\mid a)^2\\
&< &p\lambda^{-2} P_{X_0}(a)^2
= r^{-1} |\productOf(\mathcal{Y}_S^{(n)})|^{-1} \lambda^{-2}
 P_{X_0}(a)^2.
\end{eqnarray*}
Then
\[
\expect[d(V\mid\Phi)]
\le \frac{\sqrt{2^m-1}}{2\lambda}r^{-1/2},
\]
so that
\[
\expect[d(X\mid\Phi)]
\le \frac{\sqrt{2^m-1}}{2}r^{-1/2} + 1-\lambda.
\]
The proof is complete.
\end{proof}

\begin{remark}\label{directPartRemark}
The proof of Lemma~\ref{directPart} include three important ideas:
\begin{enumerate}
\renewcommand{\labelenumi}{\textnormal{\arabic{enumi})}}
\item \emph{Using $1$-random-bin maps as randomness extractors
(Definition~\ref{randomBin} with $Q=P_{\uniformDistribution{B}}$ and
$\gamma=1$).}
$1$-random-bin map is a generalization of the random-bin idea (i.e.,
the pure-random-bin map in Example~\ref{prb}) extensively used in
information theory (e.g., \cite{oohama200707,yassaee201411}) and is
similar to the idea of universal hashing \cite{cater197904}.

Our definition is stronger than the $\mathrm{universal}_2$ condition
of \cite{cater197904}, which, in probability form, only requires the
collision probability
\[
P\{F(x)=F(y)\}
\le \frac{1}{|B|}
\]
for any distinct $x,y\in A$.
However, our definition is weaker than another form of universal
hashing \cite[Definition~1.1]{mansour199301}, which requires
\[
P\{F(x)=z,F(y)=u\}
= \frac{1}{|B|^2}
\]
for any distinct $x,y\in A$ and any $z,u\in B$.

Thanks to this distinction (i.e.,
inequality~\eqref{randomBinCondition.2}), $1$-random-bin maps include
some important ensembles that have not received much attention in the
research of random number generation, such as equal-random-bin maps
(Example~\ref{erb}) and binary linear-random-bin maps
(Example~\ref{lrb}).

\item \emph{Using Jensen's inequality and the variance of output
probability of $1$-random-bin map to obtain an upper bound of
statistical distance between the distribution of $1$-random-bin
extractor output and some fixed target distribution
(Corollary~\ref{randomBinProperty.2}).}
Similar approaches have already appeared in previous research, e.g.,
\cite[Lemma~4.4]{mansour199301} (using Cauchy-Schwarz inequality),
\cite[Eq.~(32)]{oohama200707} (using Cauchy-Schwarz inequality and
Jensen's inequality),
\cite[Appendix~A]{yassaee_non-asymptotic_2013} (using Cauchy-Schwarz
inequality).

\item \emph{Utilizing the convexity of statistical distance, i.e.,
Proposition~\ref{statisticalDistanceProperty.1}.}
Simply using idea~(2) does not work for a general source that has a
subset of high-probability elements whose total probability is
negligible.
This difficulty can be overcome by partition the alphabet of source
into two parts: one part including all small-probability elements that
can effectively be handled by idea~(2), and the other part including
all remaining elements.
The success of this approach is ensured by the convexity of
statistical distance and the boundedness of statistical distance.
Similar approaches have already been used in, e.g.,
\cite{oohama200707,yassaee_non-asymptotic_2013}, both using the
triangle inequality.
\end{enumerate}

In summary, the novelty of Lemma~\ref{directPart} is a systematic
combination of the above ideas, especially idea~(1), which enables a
refined analysis of equal-random-bin maps in
Sec.~\ref{refinedAnalysis}.
\end{remark}

\begin{lemma}\label{conversePart}
Let $X=X_{\fromTo{0}{m}}$ be an $(m+1)$-tuple of random elements in
$\productOf(\mathcal{X})$ with
$\mathcal{X}=\mathcal{X}_{\fromTo{0}{m}}$.
Let
$\varphi=(\varphi_i:\mathcal{X}_i\to\mathcal{Y}_i)_{i\in\fromTo{1}{m}}$
be an $m$-tuple of extractors.
Then, for $0<r<1$,
\[
d(X\mid\varphi)
\ge  P\left\{T_X(X)\notin A_r(\mathcal{Y})\right\} - (2^m-1)r,
\]
where $d(\cdot\mid\cdot)$, $T_X$, and $A_r$ are defined by
\eqref{eq:DeltaDistance}, \eqref{eq:Statistics}, and
\eqref{eq:TypicalSets}, respectively.
\end{lemma}

\begin{proof}
Let $G=\{\omega:T_X(X)\in A_r(\mathcal{Y})\}$ and
$B=(X_0(\varphi_i(X_i))_{i\in\fromTo{1}{m}})(G^\setComplement)$.
Then by definition,
\begin{eqnarray*}
d(X\mid\varphi)
&\ge &P_{X_0(\varphi_i(X_i))_{i\in\fromTo{1}{m}}}(B)
 - P_{X_0\uniformDistribution{\productOf(\mathcal{Y})}}(B)\\
&\ge &P(G^\setComplement)
 - P_{X_0\uniformDistribution{\productOf(\mathcal{Y})}}(B).
\end{eqnarray*}
The proof will be finished if we find an upper bound of
$P_{X_0\uniformDistribution{\productOf(\mathcal{Y})}}(B)$.
It is clear that
\[
G = \bigcap_{\emptySet\ne S\subseteq\fromTo{1}{m}} C_S
\]
with $C_S=\{\omega: T_X^{S}(X)\in I_{r|\productOf(\mathcal{Y}_S)|}\}$,
so that
\[
B
= \bigcup_{\emptySet\ne S\subseteq\fromTo{1}{m}}
 (X_0(\varphi_i(X_i))_{i\in\fromTo{1}{m}})(C_S^\setComplement),
\]
and hence
\begin{eqnarray*}
P_{X_0\uniformDistribution{\productOf(\mathcal{Y})}}(B)
&\le &\sum_{\emptySet\ne S\subseteq\fromTo{1}{m}}
 P_{X_0\uniformDistribution{\productOf(\mathcal{Y})}}
 ((X_0(\varphi_i(X_i))_{i\in\fromTo{1}{m}})(C_S^\setComplement))\\
&\le &\sum_{\emptySet\ne S\subseteq\fromTo{1}{m}}
 P_{X_0\uniformDistribution{\productOf(\mathcal{Y}_S)}}
 ((X_0(\varphi_i(X_i))_{i\in S})(C_S^\setComplement))\\
&= &\sum_{\emptySet\ne S\subseteq\fromTo{1}{m}}
 \sum_{a\in\mathcal{X}_0} P_{X_0}(a)
 P_{\uniformDistribution{\productOf(\mathcal{Y}_S)}}(D_S(a))\\
&= &\sum_{\emptySet\ne S\subseteq\fromTo{1}{m}}
 \sum_{a\in\mathcal{X}_0} P_{X_0}(a)
 \frac{|D_S(a)|}{|\productOf(\mathcal{Y}_S)|}\\
&\le &(2^m-1)r,
\end{eqnarray*}
where
\[
D_S(a)
= \{(\varphi_i(X_i(\omega)))_{i\in S}:
 \omega\in C_S^\setComplement, X_0(\omega)=a\}
\]
and
\[
|D_S(a)|
\le |\{X_S(\omega): \omega\in C_S^\setComplement, X_0(\omega)=a\}|
\le r|\productOf(\mathcal{Y}_S)|
\]
because
$1/P_{X_S\mid X_0}(X_S(\omega)\mid a) = T_X^{(S)}(X(\omega))\le r|\productOf(\mathcal{Y}_S)|$ for all
$\omega\notin C_S$ with $X_0(\omega)=a$.
\end{proof}

\begin{remark}\label{conversePartRemark}
The proof of Lemma~\ref{conversePart} is mainly based on the idea of
the proof of \cite[Lemma~2.1.2]{han200300}, and is a generalization of
\cite[Lemma~2.1.2]{han200300} and \cite[Lemma~1]{bloch201006} for
uniform target distribution.
With Lemma~\ref{conversePart}, we can even obtain an exact
second-order performance, for example, in the memoryless case, which
cannot be obtained by \cite[Lemma~2.1.2]{han200300} or
\cite[Lemma~1]{bloch201006}.
\end{remark}

With the one-shot bounds established above, we can easily obtain the
achievable rate region $\region(X)$, whose proof is omitted here.

\begin{theorem}\label{regionCharacterization}
For an $(m+1)$-tuple $X$ of correlated sources (see
Definitions~\ref{problemFormulation.1}--\ref{problemFormulation.3}),
\[
\region(X)
= \region_\informationTheoretic(X)
\eqdef \{R\in\realOf{\ge 0}^m: \sumOf(R_S)\le\infEntropyRate(X_S\mid X_0)
 \text{ for all nonempty $S\subseteq\fromTo{1}{m}$}\},
\]
where
\[
\infEntropyRate(X_S\mid X_0)
\eqdef \pliminf_{n\to\infty} \frac{1}{n} \ln
 \frac{1}{P_{X_S^{(n)}\mid X_0^{(n)}}(X_S^{(n)}\mid X_0^{(n)})}
\]
is the \terminology{spectral inf-entropy rate} of $X_S$ given $X_0$.%
\footnote{
 For a sequence $(Z_n)_{n=1}^\infty$ of real-valued random variables
 $Z_n$,
 $\pliminf_{n\to\infty} Z_n
  \eqdef \sup\{r:\lim_{n\to\infty}P\{Z_n<r\}=0\}$
 is the \terminology{limit inferior in probability} of
 $(Z_n)_{n=1}^\infty$
 \cite{han_approximation_1993,han200300}.
}
\end{theorem}

Theorem~\ref{regionCharacterization} is very general.
To understand its significance, we consider its consequences in some
special cases.

\begin{corollary}\label{slepianWolfDual}
For a pair $X=(X_1^{(n)},X_2^{(n)})_{n=1}^\infty$ of correlated
stationary memoryless sources,
\[
\region(X)
= \{(R_1,R_2)\in \realOf{\ge 0}^2: R_1\le \entropy(X_1^{(1)}),
 R_2\le \entropy(X_2^{(1)}),
 R_1+R_2\le \entropy(X_1^{(1)},X_2^{(1)})\},
\]
where $\entropy(X)\eqdef \expect[-\ln P_X(X)]$ is the
\terminology{entropy} of $X$.
\end{corollary}

\begin{figure}
\centering
\includegraphics{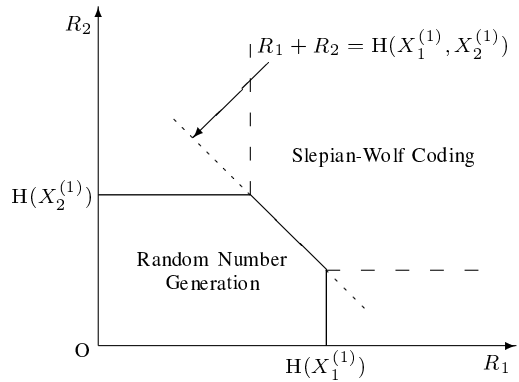}
\caption{Achievable rate regions of separate random number generation
 and Slepian-Wolf coding for correlated stationary memoryless
 sources.}
\label{fig:duality}
\end{figure}

Corollary~\ref{slepianWolfDual} provides a perfect dual of the
classical Slepian-Wolf coding theorem \cite{slepian197307}, as shown
in Fig.~\ref{fig:duality}.
This result was first discovered by Oohama in the proof of his paper
\cite{oohama200707,oohama_multiterminal_2005}.
But it was not explicitly stated, because the problem he studied only
requires that the information at each source be separately encoded and
then be jointly processed at the receiver for generating random
numbers.

The direct part of Corollary~\ref{slepianWolfDual} but with stationary
memoryless side information was also proved recently in
\cite{yassaee201411}, whose method is also valid for the direct-part
proof of correlated general sources.
Thus, the method of \cite[Theorem~1]{yassaee201411} and
\cite[Lemma~1]{bloch201006} can also be used to establish
Theorem~\ref{regionCharacterization}, though, to the author's
knowledge, Theorem~\ref{regionCharacterization} has never been
formally proved.
For the duality in this general case, the reader may compare
Theorem~\ref{regionCharacterization} with
\cite[Theorem~1]{miyake199509} or \cite[Theorem~7.2.1]{han200300}.

\begin{corollary}\label{wynerDual}
For a pair $X=(X_0^{(n)},X_1^{(n)})_{n=1}^\infty$ of correlated
sources,
\[
\region(X)
= \{R_1\in \realOf{\ge 0}: R_1\le \infEntropyRate(X_1\mid X_0)\}.
\]
\end{corollary}

Corollary~\ref{wynerDual} may be called \terminology{random number
generation with side information at the tester}, which, to some
extent, is similar to the problem of source coding with side
information at the decoder \cite{wyner197505,watanabe201504}.
A similar result, which implicitly includes Corollary~\ref{wynerDual},
was first proved in \cite{bloch201006}.

One simple application of Corollary~\ref{wynerDual} is the generation
of random seed.
When extracting randomness from a source using a seeded extractor, we
sometimes have to generate a random seed from the same source.
Loosely speaking, the source consists of two parts, $X$ and $Y$, where
$X$ is used to generate a seed in $\mathcal{S}$ and $Y$ is used to
generate random numbers.
The possible dependence between $X$ and $Y$ is however annoying.
Corollary~\ref{wynerDual} tells us that if
$n^{-1}\ln|\mathcal{S}|<\infEntropyRate(X\mid Y)$ for some $n$ large
enough, then there exists a good extractor $\varphi^{(n)}$ such that
the random seed $\varphi^{(n)}(X^{(n)})$ is not only close to uniform
but also almost independent of $Y^{(n)}$.

The implication of Corollary~\ref{wynerDual} is far-reaching, and in
particular, it is related to exposure-resilient cryptography
\cite{kamp200612}, as the following example shows.

\begin{example}\label{exposureResilient}
Let $X^{(n)}$ be an $n$-bit binary string uniformly distributed over
$\{0,1\}^n$.
Let $F^{(n)}$ be an exposure function that randomly exposes all but
$k_n=\ceil{\alpha n}$ bits of $X^{(n)}$ with $\alpha\in(0,1)$.
Exactly speaking, $F^{(n)}$ is a random map
$\{0,1\}^n\to\{0,1,\epsilon\}^n$ given by
\[
x=x_1x_2\cdots x_n
\mapsto F_1^{(n)}(x)F_2^{(n)}(x)\cdots F_n^{(n)}(x)
\]
and
\[
F_i^{(n)}(x)=
\begin{cases}
\epsilon &\text{$i\in I_n$,}\\
x_i &\text{otherwise,}
\end{cases}
\]
where $I_n$ is a random set uniformly distributed over the collection
of all $k_n$-size subsets of $\fromTo{1}{n}$.
Let $Y^{(n)}=F^{(n)}(X^{(n)})$, and then
$(X,Y)=(X^{(n)},Y^{(n)})_{n=1}^\infty$ forms a pair of correlated
sources.
Given the side information $Y^{(n)}$, it is clear that all but $k_n$
bits of $X^{(n)}$ are fixed.
Our question is how much randomness we can extract from $X$ given $Y$.
By Corollary~\ref{wynerDual}, there is a good extractor sequence
$(\varphi^{(n)}:\{0,1\}^n\to\mathcal{Y}^{(n)})_{n=1}^\infty$ for any
rate not larger than
\[
\infEntropyRate(X\mid Y) = \alpha\ln 2.
\]
A direct usage of Lemma~\ref{directPart} further shows that
\[
d(X^{(n)}\mid\varphi^{(n)})
\le \eConstant^{-n(\alpha\ln 2-\rate(\varphi^{(n)})-\smallO(1))/2}.
\]
Note that $d(X^{(n)}\mid\varphi^{(n)})$ is the average of
\(
\statisticalDistance(\varphi^{(n)}(Z^{(n)}),
 \uniformDistribution{\mathcal{Y}^{(n)}})
\)
over the set of all $n$-bit binary strings $Z^{(n)}$ in which $n-k_n$
bits are fixed and the rest are uniformly distributed, where $Z^{(n)}$
is identified with its distribution and hence the number of all
$Z^{(n)}$'s is ${n\choose k_n} 2^{n-k_n}$.
\end{example}

Owing to Theorem~\ref{regionCharacterization} (as well as its special
case \cite[Theorem~2.2.2]{han200300} for single source), the spectral
inf-entropy rate becomes the most natural measure of intrinsic
randomness (for fixed-rate randomness extraction), but unfortunately,
it does not have a simple finite-length counterpart.
A natural candidate is
\[
\infEntropy_\epsilon(X)
\eqdef \sup\left\{r:
 P\left\{\ln\frac{1}{P_X(X)}<r\right\}\le\epsilon\right\}
\]
(loosely speaking, the minimum of $-\ln P_X(X)$ with confidence level
$1-\epsilon$).
Note that $\infEntropy_0(X)$ is just the \terminology{min-entropy}
$\min\{-\ln P_X(X)\}$, which not only reveals the relation between the
min-entropy and the spectral inf-entropy rate but also explains why
the min-entropy usually underestimates the intrinsic randomness of a
source.

\section{Refined Analysis}\label{refinedAnalysis}

By Theorem~\ref{regionCharacterization} we established the complete
characterization of the achievable rate region $\region(X)$, which is
however not enough for practice because our proof is based on the
average performance of $1$-random-bin maps and hence does not give an
explicit construction of extractors.
In order to provide useful hints for designing extractors, we need a
refined analysis for some important kinds of $1$-random-bin maps,
which is the task of this section.

Markov's inequality can certainly help us relate the performance of
individual maps with the average performance of whole ensemble, but it
is too weak.
What we will use is McDiarmid's inequality \cite{mcDiarmid198900}, a
concentration inequality based on the martingale approach.

The first object of study is the pure-random-bin map.

\begin{lemma}\label{prbConcentration}
Let $X=X_{\fromTo{0}{m}}$ be an $(m+1)$-tuple of random elements in
$\productOf(\mathcal{X})$ with
$\mathcal{X}=\mathcal{X}_{\fromTo{0}{m}}$.
Let $\Phi=(\Phi_i:\mathcal{X}_i\to\mathcal{Y}_i)_{i\in\fromTo{1}{m}}$
be an $m$-tuple of pure-random-bin extractors.
Then, for $r>1$ and $s>0$,
\[
P\left\{d(X\mid\Phi) \ge
 P\{(T_X(X),\hat{T}_X(X))\notin A_r(\mathcal{Y})\times I_{rs}^m\}
 + \frac{\sqrt{2^m-1}+\sqrt{2m}}{2}r^{-1/2}\right\}
\le \eConstant^{-s},
\]
where $d(\cdot\mid\cdot)$, $T_X$, $A_r$, and $I_t$ are defined by
\eqref{eq:DeltaDistance}, \eqref{eq:Statistics},
\eqref{eq:TypicalSets}, and \eqref{eq:Interval}, respectively, and
\begin{equation}
\hat{T}_X(x)
\eqdef (\hat{T}_X^{(i)}(x))_{i\in\fromTo{1}{m}}
= \left(\frac{1}{P_{X_i}(x_i)}\right)_{i\in\fromTo{1}{m}}.
\label{eq:Statistics2}
\end{equation}
\end{lemma}

\begin{proof}
Let $\lambda=P(G)$ with
$G=\{(T_X(X),\hat{T}_X(X))\in A_r(\mathcal{Y})\times I_{rs}^m\}$.
Since the lemma holds trivially for $\lambda=0$, we suppose that
$\lambda>0$.
Then
\[
P_X= \lambda P_V+(1-\lambda) P_W,
\]
where $P_V=P_{X\mid 1_G=1}$ and $P_W=P_{X\mid 1_G=0}$.
In the same vein of the proof of Lemma~\ref{directPart}, we have
\begin{eqnarray}
d(X\mid\Phi)
&\le &\lambda d(V\mid\Phi) + 1-\lambda,
\label{eq:prbConcentrationProof:0}\\
P_V&\le &\lambda^{-1} P_X,
\label{eq:prbConcentrationProof:1}\\
\expect[d(V\mid\Phi)]
&\le &\frac{\sqrt{2^m-1}}{2\lambda}r^{-1/2}.
\label{eq:prbConcentrationProof:2}
\end{eqnarray}

Since each $\Phi_i$ is a pure-random-bin map, $d(V\mid\Phi)$ is in
fact a function of the independent random elements
\[
((\Phi_1(u))_{u\in\mathcal{X}_1},
(\Phi_2(u))_{u\in\mathcal{X}_2},
\ldots,
(\Phi_m(u))_{u\in\mathcal{X}_m})
\]
and for any $i\in\fromTo{1}{m}$ and $u\in\mathcal{X}_i$,
\[
\sup_{\substack{\varphi,\psi\in\Phi(\Omega):\\
 (\varphi_j=\psi_j)_{j\ne i}\\
 \varphi_i|_{\{u\}^\setComplement}
 =\psi_i|_{\{u\}^\setComplement}}}
 \left|d(V\mid\varphi) - d(V\mid\psi)\right|
\le P_{V_i}(u).
\]
Using McDiarmid's inequality
\cite[Lemma~1.2 and Theorem~6.7]{mcDiarmid198900},
we obtain
\[
P\{d(V\mid\Phi)-\expect[d(V\mid\Phi)]\ge\delta\}
\le \eConstant^{-2\delta^2/\chi}
\]
with
\(
\chi=\sum_{i=1}^m \sum_{u\in\mathcal{X}_i} P_{V_i}(u)^2
\)
and $\delta>0$.
Choosing $\delta=\sqrt{2s\chi}/2$, we have
\[
P\left\{d(V\mid\Phi)-\expect[d(V\mid\Phi)]
 \ge \frac{\sqrt{2s\chi}}{2}\right\}
\le \eConstant^{-s}
\]
and
\begin{eqnarray}
s\chi
&= &s\sum_{i=1}^m \sum_{u\in\mathcal{X}_i} P_{V_i}(u)^2\nonumber\\
&\le &s\sum_{i=1}^m \sum_{u\in X_i(G)} \left(
 \sum_{(a,v)\in\mathcal{X}_{\{0\}\cup \{i\}^\setComplement}}
 \frac{P_{X_0X_iX_{\{i\}^\setComplement}}(a,u,v)}{\lambda}\right)^2
 \label{eq:prbConcentrationProof:3}\\
&\le &\lambda^{-2}s\sum_{i=1}^m \sum_{u\in X_i(G)}
 P_{X_i}(u)^2\nonumber\\
&\le &m\lambda^{-2}r^{-1},
 \label{eq:prbConcentrationProof:4}
\end{eqnarray}
where \eqref{eq:prbConcentrationProof:3} follows from
$P\{V_i\in X_i(G)\}=1$ and \eqref{eq:prbConcentrationProof:1}, and
\eqref{eq:prbConcentrationProof:4} follows from a property of
elements in $X_i(G)$, that is, $1/P_{X_i}(u)>rs$ for $u\in X_i(G)$.
This implies that
\begin{eqnarray*}
P\left\{d(X\mid\Phi)\ge \lambda\expect[d(V\mid\Phi)]
 + \frac{\sqrt{2m}}{2}r^{-1/2} + 1 - \lambda\right\}
\le \eConstant^{-s},
\end{eqnarray*}
which combined with \eqref{eq:prbConcentrationProof:2} yields the
desired result.
\end{proof}

An immediate consequence of Lemma~\ref{prbConcentration} is the next
theorem.

\begin{theorem}\label{prbConcentration2}
Let $X=(X_{\fromTo{0}{m}}^{(n)})_{n=1}^\infty$ be an $(m+1)$-tuple of
correlated sources and $\Phi^{(n)}=\Phi_{\fromTo{1}{m}}^{(n)}$ an
$m$-tuple of pure-random-bin extractors such that
$(\lim_{n\to\infty}\rate(\Phi_i^{(n)}))_{i\in\fromTo{1}{m}}$ is an
interior point of $\region_\informationTheoretic(X)$.
Then, for any $\epsilon_1,\epsilon_2>0$, there exists an integer
$N=N(\epsilon_1,\epsilon_2)$ such that
\[
P\{d(X^{(n)}\mid \Phi^{(n)}) \ge \epsilon_1\}
\le \eConstant^{-\eConstant^{n(\min_{1\le i\le m}\infEntropyRate(X_i)
 -\epsilon_2)}}
\]
for all $n\ge N$.
\end{theorem}

Loosely speaking, the fraction of bad sample maps of a pure-random-bin
extractor converges to zero (as $n$ goes to infinity) in a speed of
double exponential function as long as every source to be extracted
has a nonzero spectral inf-entropy rate.
In other words, most maps are good extractors, since a pure-random-bin
map is uniformly distributed over the set of all maps with the same
domain and codomain.

\begin{example}[cf.\ \cite{kamp200612,lee201012}]
Let us resume the discussion of Example~\ref{exposureResilient}, with
also the same definitions and notations.
Lemma~\ref{prbConcentration} shows that, for $\nu>0$, the
pure-random-bin extractor
$\Phi^{(n)}:\{0,1\}^n\to\mathcal{Y}^{(n)}$ satisfies
\[
P\{\statisticalDistance(\Phi^{(n)}(Z^{(n)}),
 \uniformDistribution{\mathcal{Y}^{(n)}})
 \ge 1.21\eConstant^{-n(\alpha\ln 2-\rho_n-\nu)/2}\}
\le \eConstant^{-\eConstant^{n\rho_n}}.
\]
where $\rho_n=\max\{\rate(\Phi^{(n)}),\ln(2n)/n\}$.
Since the number of all $Z^{(n)}$'s is ${n\choose k_n} 2^{n-k_n}$, we
have
\[
{n\choose k_n} 2^{n-k_n}
 \eConstant^{-\eConstant^{n\rho_n}}
< 2^{2n} \eConstant^{-2n}
= \smallO(1),
\]
so that there is an extractor
$\varphi^{(n)}:\mathcal{X}^n\to\mathcal{Y}^{(n)}$ satisfying
\[
\statisticalDistance(\varphi^{(n)}(Z^{(n)}),
 \uniformDistribution{\mathcal{Y}^{(n)}})
= \eConstant^{-n(\alpha\ln 2-\rho_n-\smallO(1))/2}
\]
for all $Z^{(n)}$'s.
This pointwise performance is clearly much stronger than the average
performance in Example~\ref{exposureResilient}.
If using Lemma~\ref{erbConcentration} instead of
Lemma~\ref{prbConcentration}, we can further conclude that there is an
equal-bin extractor achieving this pointwise performance.
\end{example}

The next object of study is the equal-random-bin map, which has a
similar concentration phenomenon to the pure-random-bin map.

\begin{lemma}\label{erbConcentration}
Let $X=X_{\fromTo{0}{m}}$ be an $(m+1)$-tuple of random elements in
$\productOf(\mathcal{X})$ with
$\mathcal{X}=\mathcal{X}_{\fromTo{0}{m}}$.
Let $\Phi=(\Phi_i:\mathcal{X}_i\to\mathcal{Y}_i)_{i\in\fromTo{1}{m}}$
be an $m$-tuple of equal-random-bin extractors.
Then, for $r>1$ and $s>0$,
\[
P\left\{d(X\mid\Phi) \ge
 P\{(T_X(X),\hat{T}_X(X))\notin A_r(\mathcal{Y})\times I_{rs}^m\}
 + \frac{\sqrt{2^m-1}+2\sqrt{2m}}{2}r^{-1/2}\right\}
\le \eConstant^{-s},
\]
where $d(\cdot\mid\cdot)$, $T_X$, $\hat{T}_X$, $A_r$, and $I_t$ are
defined by \eqref{eq:DeltaDistance}, \eqref{eq:Statistics},
\eqref{eq:Statistics2}, \eqref{eq:TypicalSets}, and
\eqref{eq:Interval}, respectively.
If $m=1$ and $X_1$ is independent of $X_0$, then we have a slightly
improved result:
\[
P\left\{d(X\mid\Phi) \ge
 P\{(T_X(X),\hat{T}_X(X))\notin A_r(\mathcal{Y})\times I_{rs}^m\}
 + \frac{1+\sqrt{2}}{2}\eConstant^{-n\gamma/2}\right\}
\le \eConstant^{-s}.
\]
\end{lemma}

\begin{proof}
Let $\lambda=P(G)$ with
$G=\{(T_X(X),\hat{T}_X(X))\in A_r(\mathcal{Y})\times I_{rs}^m\}$.
In the same vein of the proof of Lemma~\ref{prbConcentration}, we have
\eqref{eq:prbConcentrationProof:0}--\eqref{eq:prbConcentrationProof:2}.

Since each $\Phi_i$ is an equal-random-bin map, it can be expressed as
\[
\Phi_i
= \uniformDistribution{\symmetricGroup{\mathcal{Y}_i}}
 \mapComposition\varphi_i
 \mapComposition\uniformDistribution{\symmetricGroup{\mathcal{X}_i}}
\]
where $\varphi_i$ is an arbitrary fixed equal-bin map from
$\mathcal{X}_i$ onto $\mathcal{Y}_i$.
Since the probability distribution of
$\uniformDistribution{\mathcal{Y}_i}$ is invariant under
$\uniformDistribution{\symmetricGroup{\mathcal{Y}_i}}$, we further
assume that
$\Phi_i = \varphi_i \mapComposition
 \uniformDistribution{\symmetricGroup{\mathcal{X}_i}}$.
Then $d(V\mid\Phi)$ becomes a function of the random element
$((\uniformDistribution{\symmetricGroup{\mathcal{X}_i}}(u)
 )_{u\in\mathcal{X}_i})_{i\in\fromTo{1}{m}}$.
Let
$t=(t_i:\fromTo{1}{|\mathcal{X}_i|}\to\mathcal{X}_i)
 _{i\in\fromTo{1}{m}}$
be an $m$-tuple of one-to-one maps such that
$P_{V_i}(t_i(\ell))$ is nonincreasing in $\ell$ for all
$i\in\fromTo{1}{m}$, so that 
$((\uniformDistribution{\symmetricGroup{\mathcal{X}_i}}(u)
 )_{u\in\mathcal{X}_i})_{i\in\fromTo{1}{m}}$
can be ordered as the following sequence:
\[
(
\uniformDistribution{\symmetricGroup{\mathcal{X}_1}}(t_1(1)),\ldots,
\uniformDistribution{\symmetricGroup{\mathcal{X}_1}}
 (t_1(|\mathcal{X}_1|)),\ldots,
\uniformDistribution{\symmetricGroup{\mathcal{X}_m}}(t_m(1)),\ldots,
\uniformDistribution{\symmetricGroup{\mathcal{X}_m}}
 (t_m(|\mathcal{X}_m|))
).
\]

Let
$\mathcal{Z}
 =\bigcup_{i=1}^m\{(i,j):j\in\fromTo{1}{|\mathcal{X}_i|}\}$
and $\mathcal{S}=\prod_{i=1}^m \symmetricGroup{\mathcal{X}_i}$.
For $\pi,\pi'\in\mathcal{S}$, we say $\pi=_{k,\ell}\pi'$ if
$\pi_i=\pi'_i$ for all $i<k$ and $\pi_k(t_k(j))=\pi'_k(t_k(j))$ for
all $j\le\ell$.
For $(k,\ell)\in\mathcal{Z}$ and $\pi\in\mathcal{S}$, we define
\[
f_{k,\ell}(\pi)
= \expect[d(V\mid\Phi) \mid
 (\uniformDistribution{\symmetricGroup{\mathcal{X}_i}})
 _{i\in\fromTo{1}{m}} =_{k,\ell} \pi].
\]
From Proposition~\ref{equalRandomBinProperty} it follows that
\begin{equation}
|f_{k,\ell}(\pi)-f_{k,\ell}(\pi')|
\le 2P_{V_k}(t_k(\ell))\label{eq:erbConcentrationProof:1}
\end{equation}
for all $\pi,\pi'\in\mathcal{S}$ with $\pi=_{k,\ell-1}\pi'$.
Using Proposition~\ref{mcDiarmidInequality}, we obtain
\[
P\{d(V\mid\Phi)-\expect[d(V\mid\Phi)]\ge\delta\}
\le \eConstant^{-\delta^2/(2\chi)}
\]
with
\(
\chi=\sum_{i=1}^m \sum_{u\in\mathcal{X}_i} P_{V_i}(u)^2
\)
and $\delta>0$.
Choosing $\delta=\sqrt{2s\chi}$, we have
\[
P\left\{d(V\mid\Phi)-\expect[d(V\mid\Phi)]
 \ge \sqrt{2s\chi}\right\}
\le \eConstant^{-s}
\]
and \eqref{eq:prbConcentrationProof:4} in the same vein of the proof
of Lemma~\ref{prbConcentration}.
This implies that
\begin{eqnarray*}
P\left\{d(X\mid\Phi) \ge \lambda \expect[\delta(V\mid\Phi)]
 + \sqrt{2m}r^{-1/2} + 1 - \lambda\right\}
\le \eConstant^{-s},
\end{eqnarray*}
which together with \eqref{eq:prbConcentrationProof:2} establishes the
main part of the lemma.
The last statement of the lemma can simply be proved by replacing
\eqref{eq:erbConcentrationProof:1} with
\[
|f_{1,\ell}(\pi)-f_{1,\ell}(\pi')|
\le P_{V_1}(t_1(\ell))
\]
according to the improved bound of
Proposition~\ref{equalRandomBinProperty}.
\end{proof}

An immediate consequence of Lemma~\ref{erbConcentration} is the next
theorem.

\begin{theorem}\label{erbConcentration2}
Let $X=(X_{\fromTo{0}{m}}^{(n)})_{n=1}^\infty$ be an $(m+1)$-tuple of
correlated sources and $\Phi^{(n)}=\Phi_{\fromTo{1}{m}}^{(n)}$ an
$m$-tuple of equal-random-bin extractors such that
$(\lim_{n\to\infty}\rate(\Phi_i^{(n)}))_{i\in\fromTo{1}{m}}$ is an
interior point of $\region_\informationTheoretic(X)$.
Then, for any $\epsilon_1,\epsilon_2>0$, there exists an integer
$N=N(\epsilon_1,\epsilon_2)$ such that
\[
P\{d(X^{(n)}\mid \Phi^{(n)}) \ge \epsilon_1\}
\le \eConstant^{-\eConstant^{n(\min_{1\le i\le m}\infEntropyRate(X_i)
 -\epsilon_2)}}
\]
for all $n\ge N$.
\end{theorem}

\section{An Open and Transparent Lottery Scheme for Resource
 Allocation}\label{lottery}

In today's economic systems, goods (and services) are usually
allocated by a market.
In most cases market works well, but it fails when there is a shortage
of a good (which is a ``public good'' in many cases and) whose
allocation is related to social justice issues.
Examples of such a good include traffic, fresh air, affordable
housing, education, and so on.
One feasible solution to the allocation of these resources is lottery,
operated by an authority, such as a local government.
A great advantage of lottery is its simple fairness based on
randomness, so it is widely used in many countries, e.g., affordable
housing lottery in New York \cite{newYork201308}, car license-plate
lottery in Beijing \cite{beijing201101}.

However, can we trust in the randomness of a lottery?
In particular, can we ensure that there is no fraud in the process of
a lottery?
Now we will present an open and transparent lottery scheme that is
immune to any kinds of fraud under some necessary conditions.

Suppose a public source $X$ and a lottery of $k$ participants with an
$l$-tuple $s=(s_i)_{i\in\fromTo{1}{\ell}}$ of shares $s_i$ of the
total prize (which can be a fixed amount of cash or goods), where
$1\le \ell\le k$.
$\ell$ of $k$ participants are to be chosen as the winners of prizes
$s_1,s_2,\ldots,s_\ell$, respectively, in a random order by the
lottery based on the intrinsic randomness of $X$.

To secure the process of lottery, we require that $X$ be not in the
control of anyone, or at least, that any valid fraud on $X$ cost the
cheater much more than the prize of the lottery.
A natural phenomenon that can be easily witnessed by anyone is one
kind of candidates for $X$.
Another kind of candidates is a man-made public event involving
interactions among a large number of people.
Clearly, no one can easily control such an event.
Examples of the above two kinds of sources are weather and share
prices, respectively.

The details of our lottery scheme is as follows:
\begin{enumerate}
\item Generate an electronic file $f_1$ containing an ordered list of
 all participants (numbered from $1$ to $k$) as well as all
 necessary information for identifying the participants and verifying
 their eligibility.
 Compute the hash value of $f_1$ using a cryptographic hash function
 (e.g., SHA-1) and disclose this value as well as the number $k$ to
 the public.

\item Choose and announce in public a future time interval $[t_1, t_2]$
 of $X$ for lottery.
 An estimation is needed to ensure that source $X$ outputs enough
 randomness for lottery during the chosen period.

\item Record the data of $X$ from $t_1$ to $t_2$ into an electronic
 file $f_2$.

\item Compute the lottery result by a software extracting randomness
 from $f_2$ with also a seed generated from $f_1$.
 Declare the list of winners and disclose to the public the files
 $f_1$ and $f_2$ as well as the source code of the software, so that
 everyone can verify the lottery.
\end{enumerate}

Since the statistics of $X$ may be very complicated, the success of
our scheme depends on the performance of the extractor used by the
software.
Corollary~\ref{wynerDual} and Lemma~\ref{directPart} tell us that,
whatever $X$ is, there is an ``almost-blind'' good extractor for any
rate $R$ below the information-theoretic bound
$\infEntropyRate(X\mid Y)$ of $X$ given any side information $Y$.%
\footnote{
 Lemma~\ref{directPart} shows that a $1$-random-bin extractor works
 well at any rate $R<\infEntropyRate(X\mid Y)$, regardless of the
 actual probability distribution of the correlated sources $(X,Y)$.
 Such an extractor ensemble is said to be ``blind'' or ``universal'',
 but there is no individual extractor having such a property.
 In fact, for any extractor $\varphi^{(n)}:\mathcal{X}^n\to\{0,1\}$,
 the general source
 $X=(\uniformDistribution{(\varphi^{(n)})^{-1}(y_n)})_{n=1}^\infty$
 with $y_n$ maximizing $|(\varphi^{(n)})^{-1}(y_n)|$ satisfies
 $\infEntropyRate(X)=\ln|\mathcal{X}|$ and
 $\statisticalDistance(\varphi^{(n)}(X^{(n)}),
  \uniformDistribution{\{0,1\}})=\frac{1}{2}$.
 For this reason, we only claim the existence of an ``almost-blind''
 good extractor, which is confirmed by
 Theorems~\ref{prbConcentration2} and \ref{erbConcentration2}.
}
In particular, the output of extractor is almost independent of $Y$.
Since $X$ can be regarded as the output of a virtual channel with
input $Y$, the quantity $\infEntropyRate(X\mid Y)$ actually shows the
ability of the potential adversary associated with $Y$.
This then implies that our lottery scheme is immune to any attack
based on any side information $Y$ with $\infEntropyRate(X\mid Y) > R$.

Having introduced the whole lottery scheme, we proceed to design the
lottery algorithm.
We will only present a basic framework, leaving the details to the
author's future work.

\begin{figure*}
\centering
\includegraphics{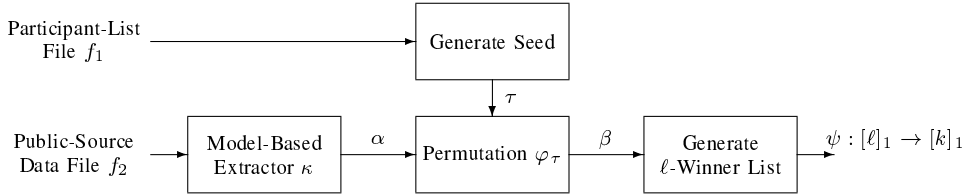}
\caption{The proposed framework of lottery algorithm.}
\label{fig:framework}
\end{figure*}

The goal of this algorithm is to extract randomness from provided data
and then to generate a uniformly distributed one-to-one map from
$\fromTo{1}{\ell}$ into $\fromTo{1}{k}$, i.e., the $\ell$-winner list
of the lottery.
Fig.~\ref{fig:framework} shows the basic idea of the algorithm, which
consists of three steps:
\begin{enumerate}
\item Process the data file $f_2$ by a model-based extractor $\kappa$,
 which is a traditional compression algorithm but with necessary
 modifications for random number generation, i.e., discarding any
 output that is possibly structured.
 Examples of this idea are \cite{zhou201204} and
 \cite{seroussi201307}.
 The output of this step is denoted by $\alpha$, a binary string.

\item Permute $\alpha$ by a permutation $\varphi_\tau$ of
 $\{0,1\}^{|\alpha|}$, which is chosen from some collection of
 permutations according to the seed $\tau$ generated from $f_1$ (by a
 hash function different from the hash function already used in the
 first step of the lottery scheme).
 The output of this step is denoted by $\beta$, also a binary string.
 If well designed, any truncation of $\varphi_\tau$ can be a good
 extractor of that rate, or equivalently, $\varphi_\tau$ is a good
 rateless extractor whose existence is confirmed by the theoretical
 performance of equal-random-bin maps
 (Theorem~\ref{erbConcentration2}).

\item Generate from $\beta$ a one-to-one map
 $\psi:\fromTo{1}{\ell}\to\fromTo{1}{k}$.

 We may, for example, use the Fisher-Yates shuffle algorithm
 \cite[Algorithm P at p.~145]{knuth199800} and the interval
 algorithm \cite{han199703} to generate from $\beta$ a random
 permutation $\pi$ of $\fromTo{1}{k}$.
 The restriction of $\pi$ to $\fromTo{1}{\ell}$ then gives a
 one-to-one map $\psi$ from $\fromTo{1}{\ell}$ into $\fromTo{1}{k}$.
 The actual data consumed for generating $\pi$ are denoted by
 $\beta'$, a truncation of $\beta$.
 Since $\varphi_\tau$ is a good rateless extractor, $\beta'$ is surely
 close to uniform (provided that data $f_2$ has enough intrinsic
 randomness).
 The length of $\beta'$ is about
 $\log_2(k!)\approx(k+\frac{1}{2})\log_2 k-k\log_2\eConstant$ bits,
 which combined with the spectral inf-entropy rate of $X$ can be used
 to determine the time interval $[t_1,t_2]$ for lottery.

 By a careful arrangement of the algorithm, we can ensure that $\psi$,
 i.e., $\pi|_{\fromTo{1}{\ell}}$ for any $\ell\in\fromTo{1}{k}$,
 always depends on the whole data $\beta'$, so that the actual
 probability distribution of $\psi$ will not deviate too much even if
 there are severe local defects in $\beta'$.%
 \footnote{
  There's no such thing as a free lunch, because we spend more
  randomness on $\psi$ than needed.
  In other words, this algorithm has an intrinsic extractor (cf.\
  Theorem~\ref{erbConcentration2}).
 }
\end{enumerate}

Let us close this section with a summary of our lottery scheme.
First and foremost, it is open and transparent, and to some extent,
immune to all kinds of fraud.
Second, it can largely reduce the cost of lottery, such as the cost
for notarization, which is sometimes a necessary part of a lottery.
Its cost-saving feature will become more prominent if it is further
standardized, e.g., using a standard open-source software as well as
standard public data sources for lottery.
Third, the proposed framework of lottery algorithm is in fact a
combination of knowledge-based approaches (Model-based extractor
$\kappa$) and non-knowledge-based approaches (Permutation
$\varphi_\tau$).
This idea is so fundamental that it is useful for any design of random
number generator.
Using a permutation as a rateless extractor is also an interesting
idea.
Finding such a permutation then becomes an important research issue.
Possible candidates include all existing encryption algorithms.

\section{Conclusion}

In this paper, we investigated the problem of separate random number
generation from correlated sources with side information at the
tester.
Tight one-shot lower and upper performance bounds, as well as the
achievable rate region, are obtained.
By a refined analysis, we further proved the doubly-exponential
concentration phenomenon of the performance of two kinds of random
extractors:
the pure-random-bin extractor and the equal-random-bin extractor.
As an application of these results, we presented an open and
transparent lottery scheme for resource allocation, which may be the
first lottery scheme immune to all kinds of fraud.

The generation of random numbers is a fundamental issue in information
science, extensively related to our understanding of randomness,
knowledge, and intelligence.
This paper, in the author's view, discovered and proved some very
fundamental facts that are necessary for the theory of universal
random number generation, which may finally be established via the
approaches of information theory, coding theory, decision theory, and
computational complexity theory.

\appendix
\section{Useful Facts}

\begin{proposition}\label{statisticalDistanceProperty.1}
If $V$, $V'$, $W$, and $W'$ are random elements in the measurable space
$(\mathcal{X}, \mathfrak{X})$, then
\[
\statisticalDistance(\lambda P_V + (1-\lambda) P_W,
 \lambda P_{V'} + (1-\lambda) P_{W'})
\le \lambda\statisticalDistance(P_V,P_{V'})
 + (1-\lambda)\statisticalDistance(P_W,P_{W'})
\]
for all $\lambda\in[0,1]$.
\end{proposition}

\begin{proof}
\begin{eqnarray*}
& &\statisticalDistance(\lambda P_V + (1-\lambda) P_W,
 \lambda P_{V'} + (1-\lambda) P_{W'})\\
& &=\sup_{A\in \mathfrak{X}} |(\lambda P_V + (1-\lambda) P_W)(A)
 - (\lambda P_{V'} + (1-\lambda) P_{W'})(A)|\\
& &=\sup_{A\in \mathfrak{X}}
 |\lambda(P_V(A)-P_{V'}(A))+(1-\lambda)(P_W(A)-P_{W'}(A))|\\
& &\le\sup_{A\in \mathfrak{X}}
 [\lambda|P_V(A)-P_{V'}(A)|+(1-\lambda)|P_W(A)-P_{W'}(A)|]\\
& &\le\lambda\statisticalDistance(P_V,P_{V'})
 + (1-\lambda)\statisticalDistance(P_W,P_{W'}).
\end{eqnarray*}
\end{proof}

\begin{proposition}\label{statisticalDistanceProperty.2}
Let $V$ and $V'$ be two random elements in the measurable space
$(\mathcal{X},\mathfrak{X})$.
Let $f$ be a measurable mapping from $(\mathcal{X},\mathfrak{X})$ to
$(\mathcal{Y},\mathfrak{Y})$.
Then
\(
\statisticalDistance(P_V,P_{V'})
\ge \statisticalDistance(P_{f(V)},P_{f(V')})
\).
\end{proposition}

\begin{proof}
\begin{eqnarray*}
\statisticalDistance(P_{f(V)},P_{f(V')})
&= &\sup_{B\in \mathfrak{Y}} |P_{f(V)}(B)-P_{f(V')}(B)|\\
&= &\sup_{B\in \mathfrak{Y}} |P_V(f^{-1}(B))-P_{V'}(f^{-1}(B))|\\
&\le &\sup_{A\in \mathfrak{X}} |P_V(A)-P_{V'}(A)|
= \statisticalDistance(P_V,P_{V'}).
\end{eqnarray*}
\end{proof}

\begin{proposition}\label{equalRandomBinProperty}
Let $\mathcal{X}=\mathcal{X}_{\fromTo{0}{m}}$ and
$\mathcal{Y}=\mathcal{Y}_{\fromTo{1}{m}}$ be two families of alphabets
with $\mathcal{X}_0$ at most countable and all the other
alphabets finite.
Let $V=V_{\fromTo{0}{m}}$ be a random element in
$\productOf(\mathcal{X})$ and
$\Phi=(\varphi_i \mapComposition \uniformDistribution{
 \symmetricGroup{\mathcal{X}_i}})_{i\in\fromTo{1}{m}}$
with $\varphi_i$ an arbitrary equal-bin map from $\mathcal{X}_i$ onto
$\mathcal{Y}_i$.

Let
$t=(t_i:\fromTo{1}{|\mathcal{X}_i|}\to\mathcal{X}_i)
 _{i\in\fromTo{1}{m}}$
be an $m$-tuple of one-to-one maps such that $P_{V_i}(t_i(\ell))$ is
nonincreasing in $\ell$ for all $i\in\fromTo{1}{m}$.
Let
$\mathcal{Z}=\bigcup_{i=1}^m\{(i,j):j\in\fromTo{1}{|\mathcal{X}_i|}\}$
and $\mathcal{S}=\prod_{i=1}^m \symmetricGroup{\mathcal{X}_i}$.
For $\pi,\pi'\in\mathcal{S}$, we say $\pi=_{k,\ell}\pi'$ if
$\pi_i=\pi'_i$ for all $i<k$ and $\pi_k(t_k(j))=\pi'_k(t_k(j))$ for
all $j\le\ell$.

For $(k,\ell)\in\mathcal{Z}$ and $\pi\in\mathcal{S}$, define
\[
f_{k,\ell}(\pi)
= \expect[d(V\mid\Phi) \mid
 (\uniformDistribution{\symmetricGroup{\mathcal{X}_i}})
 _{i\in\fromTo{1}{m}} =_{k,\ell} \pi].
\]
Then
\[
|f_{k,\ell}(\pi)-f_{k,\ell}(\pi')|\le 2P_{V_k}(t_k(\ell))
\]
for all $\pi,\pi'\in\mathcal{S}$ with $\pi=_{k,\ell-1}\pi'$.
In particular, if $P_{V_k}$ has a nonincreasing order $t_k$
independent of $V_{\{0\}\cup\{k\}^\setComplement}$, that is,
$P_{V_0V_kV_{\{k\}^\setComplement}}(a,t_k(\ell),v)$ is nonincreasing
in $\ell$ for all $a\in\mathcal{X}_0$ and
$v\in\mathcal{X}_{\{k\}^\setComplement}$, then the coefficient $2$ of
this upper bound can be replaced with $1$.
\end{proposition}

\begin{proof}
Let $\mathcal{S}'=\{\sigma\in\mathcal{S}:\sigma=_{k,\ell}\pi\}$.
Then
\begin{eqnarray}
|f_{k,\ell}(\pi)-f_{k,\ell}(\pi')|
&= & \Biggl|\sum_{\substack{\sigma\in\mathcal{S}'\\
 \sigma'=\iota\mapComposition\sigma}} \Bigl(
 P_{\uniformDistribution{\mathcal{S}'}}(\sigma)
 \statisticalDistance(
  P_{V_0((\varphi_i\mapComposition\sigma_i)(V_i))
  _{i\in\fromTo{1}{m}}},
  P_{V_0}P_{\uniformDistribution{\productOf(\mathcal{Y})}})\nonumber\\
& &- P_{\uniformDistribution{\mathcal{S}'}}(\sigma')
 \statisticalDistance(
  P_{V_0((\varphi_i\mapComposition\sigma'_i)(V_i))
   _{i\in\fromTo{1}{m}}},
  P_{V_0}P_{\uniformDistribution{\productOf(\mathcal{Y})})}
 \Bigr) \Biggr|\nonumber\\
&= & \Biggl|\sum_{\substack{\sigma\in\mathcal{S}'\\
 \sigma'=\iota\mapComposition\sigma}}
 P_{\uniformDistribution{\mathcal{S}'}}(\sigma) \Bigl(
 \statisticalDistance(
  P_{V_0((\varphi_i\mapComposition\sigma_i)(V_i))
  _{i\in\fromTo{1}{m}}},
  P_{V_0}P_{\uniformDistribution{\productOf(\mathcal{Y})}})\nonumber\\
& &- \statisticalDistance(
  P_{V_0((\varphi_i\mapComposition\sigma_i')(V_i))
  _{i\in\fromTo{1}{m}}},
  P_{V_0}P_{\uniformDistribution{\productOf(\mathcal{Y})}})
 \Bigr) \Biggr|\nonumber\\
&\le &\sum_{\substack{\sigma\in\mathcal{S}'\\
 \sigma'=\iota\mapComposition\sigma}}
 P_{\uniformDistribution{\mathcal{S}'}}(\sigma) \Bigl|
 \statisticalDistance(
  P_{V_0((\varphi_i\mapComposition\sigma_i)(V_i))_{i\in\fromTo{1}{m}}},
  P_{V_0}P_{\uniformDistribution{\productOf(\mathcal{Y})}})\nonumber\\
& &- \statisticalDistance(
  P_{V_0((\varphi_i\mapComposition\sigma_i')(V_i))
  _{i\in\fromTo{1}{m}}},
  P_{V_0}P_{\uniformDistribution{\productOf(\mathcal{Y})}})
 \Bigr|\nonumber\\
&\le &\sum_{\substack{\sigma\in\mathcal{S}'\\
 \sigma'=\iota\mapComposition\sigma}}
 P_{\uniformDistribution{\mathcal{S}'}}(\sigma) \Bigl|
 \statisticalDistance(
  P_{V_0((\varphi_i\mapComposition\sigma_i)(V_i))_{i\in\fromTo{1}{m}}},
  P_{V_0((\varphi_i\mapComposition\sigma_i')(V_i))
  _{i\in\fromTo{1}{m}}})
 \Bigr|\nonumber\\
&\le &\sum_{\substack{\sigma\in\mathcal{S}'\\
 \sigma'=\iota\mapComposition\sigma}}
 P_{\uniformDistribution{\mathcal{S}'}}(\sigma) \Bigl|
 \statisticalDistance(P_{V_0\sigma_k(V_k)V_{\{k\}^\setComplement}},
  P_{V_0\sigma_k'(V_k)V_{\{k\}^\setComplement}})
 \Bigr|\label{eq:erbPropertyProof:1}\\
&\le &P_{V_k}(t_k(\ell))+P_{V_k}(\pi_k^{-1}(\pi'_k(t_k(\ell))))
 \label{eq:erbPropertyProof:2}\\
&\le &2P_{V_k}(t_k(\ell)),\label{eq:erbPropertyProof:3}
\end{eqnarray}
where
\[
\iota
=(\id_{\mathcal{X}_1},\ldots,\id_{\mathcal{X}_{k-1}},
(\pi_k(t_k(\ell))\;\pi'_k(t_k(\ell))),
\id_{\mathcal{X}_{k+1}},\ldots,\id_{\mathcal{X}_m}),
\]
\eqref{eq:erbPropertyProof:1}
follows from Proposition~\ref{statisticalDistanceProperty.2},
\eqref{eq:erbPropertyProof:2} is a straightforward computation by
definition, and \eqref{eq:erbPropertyProof:3} follows from the
nonincreasing-probability property of $t_k$.
The proof is complete by noting that if $P_{V_k}$ has a nonincreasing
order $t_k$ independent of $V_{\{0\}\cup\{k\}^\setComplement}$, then
step \eqref{eq:erbPropertyProof:2} can be replaced with
\[
P_{V_k}(t_k(\ell))-P_{V_k}(\pi_k^{-1}(\pi'_k(t_k(\ell)))),
\]
so that the last upper bound becomes $P_{V_k}(t_k(\ell))$.
\end{proof}

\begin{proposition}%
[A modified version of
{\cite[Corollary~6.10]{mcDiarmid198900}}]\label{mcDiarmidInequality}
Let $Z=(Z_i)_{i=1}^n$ be an $n$-tuple of random variables in
$\productOf(A)$ with $A=(A_i)_{i=1}^n$.
Let $D_i$ be the corresponding subset of
$\productOf(A_{\fromTo{1}{i}})$ such that
$P_{Z_{\fromTo{1}{i}}}(D_i^\setComplement)=0$.
Let $f:\productOf(A)\to\real$ be a measureable function such that
$f(Z)$ is integrable.
Suppose that there are constants $c_1,\ldots,c_n$ so that
\[
\left|
 \expect\left[f(Z)\mid
 Z_{\fromTo{1}{k}}=z_{\fromTo{1}{k}}\right]
 -\expect\left[f(Z)\mid
 Z_{\fromTo{1}{k}}=z'_{\fromTo{1}{k}}\right]
 \right|
\le c_k
\]
for each $k\in\fromTo{1}{n}$ and any
$z_{\fromTo{1}{k}},z'_{\fromTo{1}{k}}\in D_k$ satisfying
$z_{\fromTo{1}{k-1}}=z'_{\fromTo{1}{k-1}}$.
Then for any $t>0$,
\[
P\left\{\left|f(Z)-\expect f(Z)\right|\ge t\right\}
\le 2\eConstant^{-2t^2/\sum_{i=1}^n c_i^2}.
\]
\end{proposition}

\begin{proof}
See \cite[Corollary~6.10]{mcDiarmid198900}.
\end{proof}

\omitted{%
\section{Omitted Stuff}

\begin{proof}[Proof of Theorem~\ref{regionCharacterization}]
1) Direct part:
Suppose that $R\in \region_\informationTheoretic(X)$.
For $\gamma\in(0,\frac{1}{2}\min_{i:R_i>0}R_i)$, we define
\[
\mathcal{Y}_i^{(n)}=\fromTo{1}{\eConstant^{n\max\{R_i-2\gamma,0\}}}
\]
and use $1$-random-bin maps (see Examples~\ref{prb}--\ref{lrb})
as the $m$-tuple $\Phi^{(n)}$ of random extractors.
Lemma~\ref{directPart} with $r=\eConstant^{n\gamma}$ shows that
\[
\expect[d(X^{(n)}\mid\Phi^{(n)})]
\le P\{T_{X^{(n)}}(X^{(n)})\notin A_r(\mathcal{Y}^{(n)})\}
 + \frac{\sqrt{2^m-1}}{2}\eConstant^{-n\gamma/2},
\]
so that
\[
\lim_{n\to\infty} \expect[d(X^{(n)}\mid\Phi^{(n)})]
\le \lim_{n\to\infty}
 P\{T_{X^{(n)}}(X^{(n)})\notin A_r(\mathcal{Y}^{(n)})\}
= 0
\]
by the definitions of $\region_\informationTheoretic(X)$ and
$\infEntropyRate(X_S|X_0)$.
This implies that there is an $m$-tuple $\varphi^{(n)}$ of extractors
such that
$\rate(\varphi_i^{(n)})=n^{-1}\ln\floor{e^{n\max\{R_i-2\gamma,0\}}}$
for all $i\in\fromTo{1}{m}$ and
$\lim_{n\to\infty} d(X^{(n)}\mid\varphi^{(n)}) = 0$.

In order to complete the proof, we use the diagonal line argument.
Choose a decreasing sequence $(\gamma_k)_{k=1}^\infty$ converging to
zero.
For each $\gamma_k$, we repeat the above argument to obtain an
$m$-tuple of extractors, denoted $\varphi^{(n,k)}$, and we denote
by $N(k)$ the least integer such that
$d(X^{(n)}\mid\varphi^{(n,k)})\le \gamma_k$
for all $n\ge N(k)$.
Define the new $m$-tuple $\psi^{(n)}$ of extractors by
\[
\psi^{(n)}=\varphi^{(n,\max(\{1\}\cup\{k:N(k)\le n\}))}.
\]
Then
\[
\liminf_{n\to\infty} \rate(\psi_i^{(n)})
\ge \sup_{k}\lim_{n\to\infty}\rate(\varphi_i^{(n,k)})
= R_i
\quad\text{for all $i\in\fromTo{1}{m}$}
\]
and
\[
\limsup_{n\to\infty} d(X^{(n)}\mid\psi^{(n)})
\le \inf_{k} \gamma_k
= 0.
\]
Therefore, $R\in \region(X)$.

2) Converse part:
Suppose that $R\in \region(X)$.
Then there exists an $m$-tuple $\varphi^{(n)}$ of extractors such that
$\liminf_{n\to\infty} \rate(\varphi_i^{(n)})\ge R_i$ for all
$i\in\fromTo{1}{m}$ and
$\lim_{n\to\infty} d(X^{(n)}\mid\varphi^{(n)})=0$.
By Lemma~\ref{conversePart} with $r=e^{-n\gamma}$,
\[
\liminf_{n\to\infty} P\{T_{X^{(n)}}(X^{(n)})\in A_r(\mathcal{Y}^{(n)})\}
\ge \liminf_{n\to\infty}
 [1-(2^m-1)\eConstant^{-n\gamma}-d(X^{(n)}\mid\varphi^{(n)})]
= 1
\]
for any $\gamma>0$.
This implies that, for every nonempty $S\subseteq\fromTo{1}{m}$ and
any $\gamma>0$,
\begin{eqnarray*}
\sumOf(R_S)
&\le &\sum_{i\in S} \liminf_{n\to\infty} \rate(\varphi_i^{(n)})
\le \liminf_{n\to\infty} \sum_{i\in S} \rate(\varphi_i^{(n)})\\
&= &\liminf_{n\to\infty} \frac{1}{n}
 \ln|\productOf(\mathcal{Y}_S^{(n)})|
< \infEntropyRate(X_S\mid X_0)+\gamma,
\end{eqnarray*}
so that $\sumOf(R_S)\le\infEntropyRate(X_S\mid X_0)$, and therefore
$R\in \region_\informationTheoretic(X)$.
\end{proof}%

}

\small
\bibliographystyle{IEEEtran}
\bibliography{IEEEfull,urng}

\end{document}